# Superconducting and Structural Transitions in the β-Pyrochlore Oxide KOs$_2$O$_6$ under High Pressure

Hiroki OGUSU, Nao TAKESHITA[1], Koichi IZAWA[2], Jun-ichi YAMAURA, Yasuo OHISHI[3], Satoshi TSUTSUI[3], Yoshihiko OKAMOTO, and Zenji HIROI [*]

*Institute for Solid State Physics, University of Tokyo, 5-1-5 Kashiwanoha, Kashiwa, Chiba 277-8581, Japan*
[1]*Nanoelectronics Research Institute, National Institute of Advanced Industrial Science and Technology, Tsukuba, Ibaraki 305-8562, Japan*
[2]*Department of Physics, Tokyo Institute of Technology, Meguro, Tokyo 152-8551, Japan*
[3]*Japan Synchrotron Radiation Research Institute, SPring-8, Sayo, Hyogo 679-5198, Japan*

Rattling-induced superconductivity in the β-pyrochlore oxide KOs$_2$O$_6$ is investigated under high pressure up to 5 GPa. Resistivity measurements in a high-quality single crystal reveal a gradual decrease in the superconducting transition temperature $T_c$ from 9.7 K at 1.0 GPa to 6.5 K at 3.5 GPa, followed by a sudden drop to 3.3 K at 3.6 GPa. Powder X-ray diffraction experiments show a structural transition from cubic to monoclinic or triclinic at a similar pressure. The sudden drop in $T_c$ is ascribed to this structural transition, by which an enhancement in $T_c$ due to a strong electron-rattler interaction present in the low-pressure cubic phase is abrogated as the rattling of the K ion is completely suppressed or weakened in the high-pressure phase of reduced symmetry. In addition, we find two anomalies in the temperature dependence of resistivity in the low-pressure phase, which may be due to subtle changes in rattling vibration.

[*]E-mail address: hiroi@issp.u-tokyo.ac.jp

## 1. Introduction

The β-pyrochlore oxides AOs$_2$O$_6$ have recently been attracting much attention,[1-4] as they exhibit an unusual atomic vibration in a solid called rattling,[5,6] which is essentially a local, anharmonic vibration of a heavy ion confined in an oversized atomic cage. They crystallize in a cubic crystal structure of space group $Fd\bar{3}m$ with all the atoms sitting at special positions, i.e., the A, Os and O atoms at the 8b, 16c and 48f sites, respectively.[7] The A atom is located in the $T_d$ site symmetry and surrounded by 6 nearest- and 12 next-nearest-neighbor oxide atoms, while OsO$_6$ octahedra are connected to each other by vertices to form a three-dimensional framework. A virtual size mismatch between the guest A ion and the cage made of the octahedra allows the guest ion to move almost freely with an unusually large excursion inside the cage.[8] Evidence of rattling in β-pyrochlore oxides has been obtained from structural analyses showing large atomic displacement parameters[7,9] or heat capacity and spectroscopic measurements that find Einstein-like modes with low energies of 2-7 meV.[2,3,10-14]

The most intriguing issue on rattling is to understand its effect on the electronic properties of materials. It is considered that, in β-pyrochlores, the large resistivity and its anomalous temperature dependence with a concave-downward curvature in a wide temperature range are due to a strong scattering of electrons by rattling.[15] Moreover, the increase observed in the spin-lattice relaxation rate of A-nucleus NMR arises from a strong electron-lattice coupling of the same origin.[15,16] As a result of this large electron-rattler interaction, β-pyrochore oxides undergo superconducting transitions at relatively high temperatures of $T_c$ = 9.6, 6.3, and 3.3 K for A = K, Rb, and Cs, respectively. Nagao *et al.* suggested that the superconductivity is really induced by the rattling itself, because the estimated average frequency of phonons mediating Cooper pairing coincides with the energy of the rattling for each of the three compounds.[4]

Unique chemical trends for various parameters are observed in the series of β-pyrochlore oxides: the rattling intensity, the extent of electron-rattler interaction, and $T_c$ increase systematically from Cs to K.[17] Particularly interesting is the fact that the superconductivity changes its character from weak coupling to extremely strong coupling toward K.[3] Since a previous structural study revealed that the size of the cage remains almost intact among the three compounds,[7] these variations should be ascribed to the increase in the guest-free space, as the ionic radius of the A ion decreases markedly from Cs to K. Thus, the guest-free space is a key parameter for adjusting the rattling intensity. One experimental method of tuning the guest-free space systematically is to chemically mix two A elements in a crystal. However, this must cause certain randomness that might mask intrinsic properties. In contrast, squeezing the compound under high pressure would give a better opportunity to study the relation between the guest-free space and the rattling or the electronic properties of β-pyrochlore oxides.

A few high-pressure (HP) experiments have already been carried out using polycrystalline samples of β-pyrochlore oxides. Muramatsu *et al.* measured resistivity in a cubic-anvil cell filled with Fluorinert under HP up to 12 GPa and found that, common to the three compounds, $T_c$ increases initially with pressure, saturates, and then decreases to vanish above a critical pressure, resulting in a domelike pressure dependence of $T_c$;[18] The critical pressures were approximately 6, 7, and 12 GPa for K, Rb, and Cs, respectively. On the other hand, Miyoshi *et al.* found, in their magnetization measurements using a diamond-anvil cell (DAC) with Daphne 7373 oil under HP up to 10 GPa, similar $T_c$ domes for K and Rb, but a saturating behavior at 8.8 K for Cs instead.[19] On the other hand, electronic structure calculations have shown that the density of states decreases gradually and only slightly with increasing pressure.[20] Therefore, the observed complicated pressure dependence of $T_c$ is not understandable in the framework of the simple BCS theory and has not yet been explained properly. Note, however, that experimental results can be markedly different between polycrystalline and single-crystal samples in the case of β-pyrochlore oxides.[4] In order to collect reliable data on the pressure dependence of $T_c$, further HP experiment using a single crystal is necessary. Very recently, Isono et al. carried out HP heat capacity measurements in a DAC filled with Ar on single crystals of K and Rb and found abrupt vanishments of $T_c$ at 5.2 and 6.0 GPa, respectively.[21] On one hand, Katrych et al. performed HP X-ray diffraction (XRD) experiments at room temperature (RT) on a powder sample of $KOs_2O_6$ and showed that the cubic pyrochlore structure is robust under compression up to 32.5 GPa.[22] They also found a new triclinic phase at $P = 3$ GPa and $T = 1173$ K, which was substantially different from the pyrochlore structure, containing edge-sharing $OsO_6$ octahedra in addition to original corner-sharing ones.

In this study, we have carried out resistivity measurements on a high-quality single crystal of $KOs_2O_6$ under HP up to 5 GPa. In addition, powder XRD experiments at 10 K under HP have been performed to explore the possibility of a structural transition at low temperatures. A reliable pressure dependence of $T_c$, which is substantially different from that previously reported, is obtained. We find that $T_c$ drops suddenly at a critical pressure of 3.6 GPa, where probably a structural transition to a lower symmetry takes place so as to reduce or suppress the rattling vibrations of the K atom.

## 2. Experimental

Single crystals of $KOs_2O_6$ were prepared by the flux method starting from a 1.3:1 mixture of $KOsO_4$ and $OsO_2$ in a quartz ampoule at 748 K for 24 h. Polycrystalline samples were prepared as reported previously.[1] Resistivity measurements were carried out on a single crystal of $0.3 \times 0.1 \times 0.1$ mm$^3$ size by the four-probe method at a current of 10 mA. High pressures from 1.0 to 5.0 GPa were applied to the sample during the measurements in a cubic-anvil-press apparatus comprising three pairs of counteranvils made of sintered diamond.[23] Daphne 7474 oil was used as a pressure-transmitting medium; it remains liquid below 3.6 GPa at RT and may guarantee good hydrostatic compression.[24] An isobaric measurement, enabled by applying a constant load on the anvil cell, was performed upon heating from 3 to 300 K. Pressure for the next run was always increased at RT, where the pressure medium remained in liquid state or soft enough to generate a uniform, hydrostatic pressure around the sample. Actual pressure exerted on the sample was estimated by measuring the changes in resistivity associated with the structural phase transitions of Bi at 2.55, 2.7, and 7.7 GPa, of Te at 4.0 GPa, and of Sn at 9.4 GPa in different runs.[23] Powder XRD data were collected at 10 K using synchrotron radiation with a wavelength of 41.222 pm in the BL-10XU beam line at the SPring-8 facility. A polycrystalline sample was put in a standard DAC together with liquid Ar as a pressure-transimitting medium. Pressure was applied at 10 K, not at RT, in order to avoid Ar from being absorbed by the sample, although this might cause a certain inhomogeneous pressure distribution. Pressure calibration was done by recording a change in the frequency of the fluorescence from a Ruby crystal present near the sample in the DAC cell.

## 3. Results

Figure 1 shows nineteen sets of isobaric resistivity $r$ data measured between $P = 1.0$ and 5.0 GPa on a single crystal of $KOs_2O_6$. The $\rho$ at 1.0 GPa resembles that reported on a different crystal at ambient pressure,[3] showing a nearly equal concave-downward curvature in the whole temperature range above $T_c$. In contrast, a small upturn emerges in the 1.3 GPa data at low temperatures, as shown in the inset: $\rho$ begins to shift upward below 14 K from a curve expected by high-temperature extrapolation. Moreover, two similar anomalies are observed at 16 and 23 K in the 1.6 GPa data. As pressure further increases, they move to higher temperatures and then disappear above 3.6 GPa. The single anomaly observed at 1.3 GPa may correspond to two anomalies that occur at nearly equal temperatures. On the other hand, above 3.7 GPa, another anomaly is detected at a higher temperature, again moving to higher temperatures with increasing pressure. For example, at the maximum pressure of 5.0 GPa, the temperature derivative of the resistivity curve clearly changes its sign at 205 K. We call the temperatures of these three anomalies from the low-temperature side $T_{O1}$, $T_{O2}$, and $T_s$. They have never been observed in previous resistivity measurements in polycrystalline samples.[18]

The isothermal resistivity always shows a positive pressure dependence, which is markedly large especially at low temperatures. As a result, as pressure increases, the overall temperature dependence changes from metallic behavior to become almost flat, followed by a steep downward shift below 50 K. Moreover, the normal-state resistivity just above $T_c$ increases with pressure, as shown in Fig. 1(b). Similar pressure dependences have been observed in other β-pyrochlores[18] as well as in the α-pyrochlore oxide $Cd_2Re_2O_7$.[25] This common feature is probably related to the fact that all these pyrochlore oxides are semimetals with electron and hole bands:[8,26,27] the application of pressure may reduce the overlap between them and make the carrier density small and the Fermi energy low, which cause such a flat temperature dependence of resistivity at high temperatures above the Fermi temperature and an increase in residual resistivity. An abrupt upward shift in resistivity is observed between 3.6 and 3.7 GPa in Fig. 1(a), which may be due to the solidification of Daphne 7474 oil.[24]

The pressure dependence of the superconducting transition is shown in Fig. 1(b). For $P = 1.0$ to 3.5 GPa, a sharp drop is observed within a transition width range of 0.1-0.3 K, as sharp as observed at ambient pressure.[3] In general, a broad superconducting transition is often observed in HP experiments owing to certain inhomogeneity in a sample or a pressure distribution inside a HP cell. It was in fact the case for previous HP experiments on KOs$_2$O$_6$, where the transition width became more than 1 K at 4 GPa.[18] Thus, the sharp transition observed in the present study indicates better sample quality and a more uniform pressure distribution attained. The $T_c$ defined here as a zero-resistive temperature increases slightly from 9.60 K at ambient pressure up to 9.75 K at 1.0 GPa, decreases gradually with increasing pressure, and then reached to 6.55 K at 3.5 GPa.

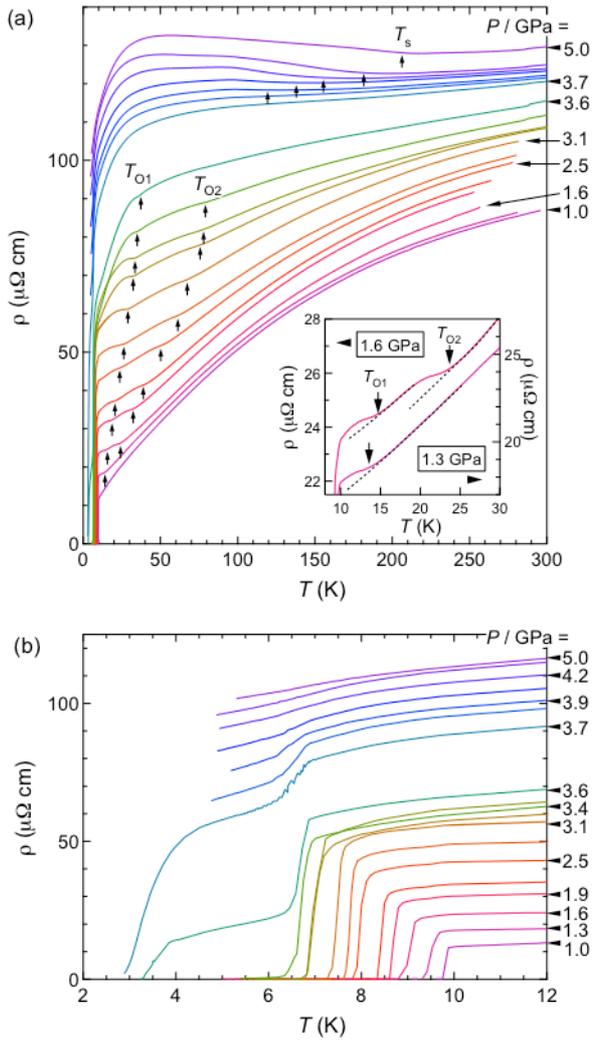

Fig. 1. (Color Online) Isobaric resistivity of a single crystal of β-KOs$_2$O$_6$ measured on heating under various pressures. The applied pressures are 1.0, 1.3, 1.6, 1.9, 2.2, 2.5, 2.8, 3.1, 3.3, 3.4, 3.5, 3.6, 3.7, 3.8, 3.9, 4.0, 4.2, 4.5, and 5.0 GPa from bottom to top. The overall temperature dependence is shown in (a), and the low-temperature part is expanded in (b), showing superconducting transitions. The inset in (a) expands around two anomalies, named $T_{O1}$ and $T_{O2}$, for the 1.3 and 1.6 GPa data.

Surprisingly, increasing pressure by only 0.1 GPa results in a completely different behavior: a two-step transition appears at 3.6 and 3.7 GPa. The first drop occurs at approximately 6.5 K at both pressures, the same as that at $T_c$ at 3.5 GPa, while the second drop to $\rho = 0$ occurs at 3.2 and 2.9 K at 3.6 and 3.7 GPa, respectively. We observed a distinct current dependence of resistivity of the temperature range between the two drops: the first drop tended to dis-appear with increasing current density. Therefore, the high-temperature transition is not bulk in nature and must be due to the formation of a filamentary superconducting path. This means that two superconducting phases with different $T_c$s coexist in the pressure range above 3.6 GPa approximately below 4.0 GPa. It is plausible that there is a first-order phase transition that accompanies a two-phase equilibrium region. We call this lower critical pressure $P_s$.

Figure 2 shows a comparison of the pressure dependences of $T_c$ obtained in the present study with those obtained previously. Below $P_s$, all the three datasets coincide with each other. Above $P_s$, $T_c$ = 5.7 K at 4 GPa was reported by Muramatsu et al.[18] However, since they determined this at the midpoint of a broadened transition, the actual $T_c$ should have been lower. On the other hand, a diamagnetic response in magnetization measurements by Miyoshi et al. had already become obscured at 3.5 GPa, and no data above that was given.[19] Thus, our observation of the sudden drop in $T_c$ at $P_s$ is in line with the previous data, and has been made owing to improved experimental conditions. In contrast, the pressure dependence of $T_c$ reported by Isono et al. in their heat capacity measurements in a single crystal of KOs$_2$O$_6$ is considerably different from the above data: their $T_c$ decreases gradually with pressure, but is still 6.0 K at 5.0 GPa, and suddenly vanishes above 5.2 GPa.[21] This vanishment may correspond to the drop in $T_c$ at $P_s$ in our experiments, although the two critical pressures are quite different. We do not know the reason for this difference, but speculate that the transition can be affected seriously by the nature of pressure that depends on the equipment used and, more importantly, on the chosen pressure medium. It is generally known that some pressure-induced transitions are very sensitive to the choice of pressure medium. For example, the α-to-ω transition in titanium metal changes from 4.9 to 10.5 GPa depending on the pressure medium.[28]

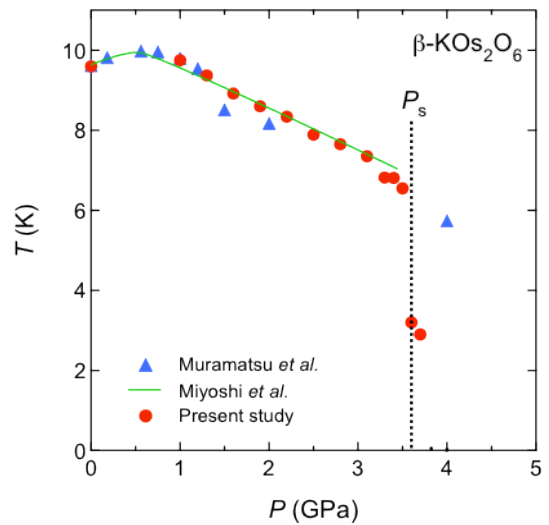

Fig. 2. (Color Online) Pressure dependence of $T_c$. The present data (circle) are compared with previous data given by Muramatsu et al. (triangle)[18] and Miyoshi et al. (solid line).[19] $T_c$ drops suddenly at $P_s$ in the present study.

X-ray diffraction experiments demonstrate that this transition is possibly related to a pressure-induced structural transition. Figure 3 shows three XRD patterns taken at $T = 10$ K and $P = 0.9$, 3.9, and 5.2 GPa. The pattern taken at 0.9 GPa resembles that at ambient pressure[1]) with all the peaks indexed by a face-centered cubic cell except for several weak peaks from a small amount of impurity phases such as Os and $OsO_2$. At 3.9 GPa, the major diffraction peaks remain unchanged but are broadened considerably, suggesting that there is an inevitable distribution in pressure inside the DAC; Ar has already been solidified at this pressure and temperature, as evidenced by the emregence of a new peak near $2\theta = 9°$ from crystalline Ar. In contrast, at 5.2 GPa, each of the three major peaks of 111, 311, and 222 indices splits into at least two peaks. This implies that the crystal system has changed from cubic to monoclinic or triclinic. This HP form is different from that reported by Katrych *et al.* at high pressures and high temperatures,[22]) because a powder diffraction pattern calculated based on the crystal structure they gave is completely dissimilar from the observed one in Fig. 3. Our HP phase must be a slight modification of the original cubic structure made of only corner-sharing $OsO_6$ octahedra. Moreover, note that Katrych *et al.* observed no change in the XRD pattern up to 32.5 GPa at RT, not at a temperature as low as 10 K in our experiments.

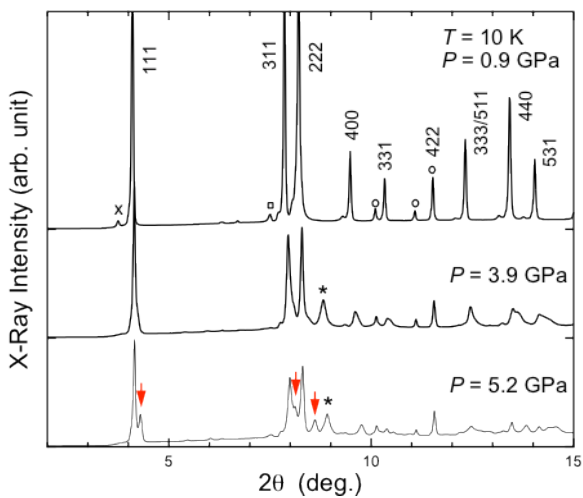

Fig. 3. (Color Online) Powder X-ray diffraction patterns taken at $T = 10$ K and $P = 0.9$ (top), 3.9 (middle), and 5.2 GPa (bottom). A synchrotron radiation with a wavelength of 41.222 pm was used in the BL-10XU beamline at the SPring-8 facility. Peaks with indices in the top pattern are from a cubic β-pyrochlore phase. Weak peaks marked by a square and circles correspond to impurity phases of $OsO_2$ and Os, respectively. The peak marked by a cross is unknown. The peaks marked by asterisks in the middle and bottom patterns correspond to solidified Ar put as a pressure-transmitting medium, which is not observed at 0.9 GPa because it may be broad and overlap with the intense 222 peak from the sample. The peaks marked by arrows emerge as a result of a structural transition to a monoclinic or triclinic structure.

The critical pressure for the structural transition may exist between 3.9 and 5.2 GPa. We think that this structural transition corresponds to the transition observed at $P_s$ in the resistivity measurements, although the two critical pressures are somewhat different. This difference may also be reconciled, however, if one takes into account the differences in the samples used (single crystals or polycrystals), experimental setups (cubic- or diamond-anvil cells), and pressure media. In addition, had there been another transition at 3.9-5.2 GPa, we should have observed a corresponding change in our resistivity data at such pressures. As shown in Fig. 1(a), however, the data between 3.7 and 5.0 GPa varies smoothly with pressure without any discernible anomaly.

From a combination of resistivity and structural investigations under HP, we conclude that a first-order structural transition takes place at $P_s = 3.6$ GPa, probably from cubic to monoclinic or triclinic symmetry. The low-pressure (LP) phase takes high $T_c$s of 9.75-6.55 K, while the HP phase has low $T_c$s below 3.3 K. The two-step transition observed just above $P_s$ in Fig. 1(b) is apparently due to the coexistence of these two phases, as expected from the first-order nature of the transition. Figure 4 shows a $P$-$T$ phase diagram for β-$KOs_2O_6$. $T_c$ increases slightly from 9.60 K at ambient pressure to 9.75 K at 1.0 GPa and then gradually decreases with pressure. Above 1.3 GPa, two anomalies, $T_{O1}$ and $T_{O2}$, appear and move to higher temperatures with increasing pressure. $T_c$ seems not affected by these anomalies, but decreases gradually with $P$. Note, however, that the $T_{O1}$ and $T_{O2}$ lines are merged around the top of the $T_c$ dome and disappear or hidden by the superconducting transition, suggesting a certain relationship between these temperatures. On the other hand, at $P_s$, $T_c$ suddenly becomes almost half, and the two anomalies disappear. The third anomaly $T_s$ shows up above $P_s$, but this may correspond to a phase boundary between the LP and HP phases. Actually, the fact that Katrych *et al.* observed no transitions at RT[22]) means that the boundary is located below RT even at 32.5 GPa. As shown in Fig. 4, we call the high-temperature cubic phase of space group $Fd$-$3m$ phase I, the intermediate ones below the $T_{O1}$ and $T_{O2}$ lines phases II and II', respectively, and the HP monoclinic or triclinic one phase III.

## 4. Discussion

Let us discuss what is happening under high pressure. It is considered that a rattling-induced, strong-coupling superconductivity occurs in β-$KOs_2O_6$ through a strong electron-rattler interaction coming from the anharmonicity of rattling.[4]) One simple idea to explain the observed gradual decrease in $T_c$ with pressure is that compressing the cage suppresses the rattling intensity and thus reduces the extent of electron-rattler interaction. The rattling intensity in β-pyrochlores can be measured using the guest-free space $d_{gfs}$, which is the distance obtained by subtracting the ionic radii of an oxide ion (140 pm) and an A ion (138 pm for $K^+$) from the A-O bond length determined by structural refinement: the larger the $d_{gfs}$, the more intense the rattling.[7]) The rattling may terminate when $d_{gfs}$ becomes zero, where the A ion can fit the cage and becomes an harmonic oscillator, as in conventional oxide compounds. Using the coordinate parameter $x$ for the 48f oxide atom, which is the only coordinate parameter in the structure, $d_{gfs}$ is equal to $a(0.625 - x) - 140 - 138$. The lattice constant $a$ decreases from 1010.1 pm at ambient pressure to 1001.2 pm at $P_s$, reduced by only 0.9%.[7]) Although the $x$ at HP is not available at the moment, it may be reasonable to assume that it does not change much with pressure. $x$ is always close to 0.3125 or slightly larger in most pyrochlore oxides.[29] For example, $x$ is equal to 0.3145 for β-$KOs_2O_6$ and 0.3137 for α-$Cd_2Re_2O_7$ at RT.[7,17] Since the coordination around a transition metal ion becomes a regular octahedron for $x$ = 0.3125, $x$ cannot deviate much from this value unless a large deformation of the octahedron is allowed, which must be unfavorable in terms of electronic energy as well as of Madelung energy. Assuming $x$ = 0.3145 under pressure gives a change in $d_{gfs}$ by just 8% from 36 to 33 pm toward $P_s$. This change is obviously too small to say that rattling is suppressed owing to the compression of the cage. On the other hand, large initial increases in $T_c$ have been observed for $RbOs_2O_6$ and $CsOs_2O_6$ under pressure, which may imply that the electron-rattler interaction is rather enhanced by pressure. Thus, we have not yet reached a reasonable explanation for the observed complicated pressure dependences of $T_c$ in the cubic phase. A more systematic HP study, particularly on the structure, using high-quality single crystals of all the three compounds is required for a further understanding.

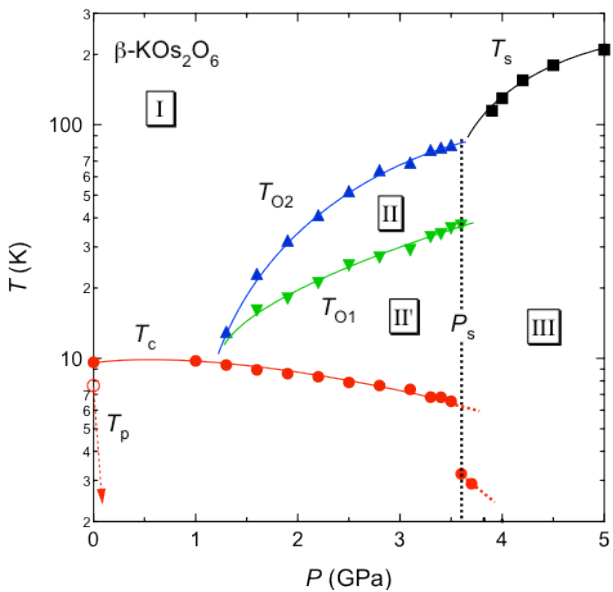

Fig. 4. (Color Online) Pressure-temperature phase diagram for β-$KOs_2O_6$. Phase I is a high-temperature, low-pressure phase crystallizing in the cubic pyrochlore structure of the space group $Fd$-$3m$, where an intense rattling of the K ion has been observed. Phase III possesses a monoclinic or triclinic structure with lower $T_c$s. A phase boundary between them exists at $P_s$ = 3.6 GPa, which is probably followed by the $T_s$ line at high temperatures. Phase II appears below $T_{O2}$ and is replaced by phase II' below $T_{O1}$. $T_p$, a critical pressure for a first-order structural transition called the rattling transition, is completely suppressed below 0.02 GPa.[35]

The high symmetry of the surrounding cage is important for the rattling phenomena; in a cage with low symmetry, a rattler tends to be trapped at off-center positions with a lower potential energy, particularly at low temperatures with weaker thermal vibrations. The $T_d$ symmetry preserved for the cage in β-pyrochlores must be crucial to the intense rattling of A ions around the on-center position. The structural transition at $P_s$ must break the $T_d$ symmetry and stop the rattling completely or at least weaken it significantly. This causes the large reduction in $T_c$ observed at $P_s$, because a certain energy gain in the Cooper pairing due to the rattling has been removed. From another perspective, $KOs_2O_6$ is a superconductor even without the rattling, if the rattling has stopped completely above $P_s$; the rattling only enhances $T_c$. Hattori and Tsunetsugu have theoretically analyzed the mechanism of superconductivity in β-pyrochlore oxides.[30] They found that two kinds of phonons are necessary in order to reproduce the chemical trend of $T_c$: one is a rattling mode with a low energy and the other is a normal Debye mode with a high energy at 260 K. Adopting appropriate parameters, they estimate $T_c$ to be 6.5 K only from the contribution of a Debye phonon and to increase to 10.5 K by adding a rattling phonon. Thus, there is an enhancement in $T_c$ by 4 K due to the rattling at ambient pressure.[30] The observed drop in $T_c$ by 3.2 K at $P_s$ is comparable to the theoretical expectation, if the contribution of the rattling has completely vanished above $P_s$.

The two weak anomalies at $T_{O1}$ and $T_{O2}$ can be attributed to other weak structural transitions concerning rattling vibration. It is clear that, in β-pyrochlores, the magnitude and temperature dependence of resistivity are dominantly governed by the electron scattering due to the rattling. Thus, any transition that affects resistivity much should be related to a change in rattling vibration. In $KOs_2O_6$, there is a weak first-order structural transition at $T_p$ = 7.6 K below $T_c$ at ambient pressure,[3,31] which is considered to be a liquid-gas transition concerning the rattling degree of freedom and is called the rattling transition.[32-34] The lattice constant is increased by only 0.01% below $T_p$,[33] so that pressure should decrease $T_p$. In fact, recent specific heat measurements by Umeo *et al.* found that the transition vanished even at only 0.02 GPa.[35] Thus, the present two anomalies are not directly related to the rattling transition. We speculate that they evidence successive, second-order structural transitions that involve small structural changes and weak reductions in symmetry, much weaker than those at $P_s$; either phases II or II' may belong to one of the maximal subgroups of $Fd$-$3m$. The XRD data at 0.9 and 3.9 GPa shown in Fig. 3 should show diffraction patterns for phases I and II', respectively. However, it is difficult to see a small difference between them, if any, owing to the marked peak broadening in the 3.9 GPa data. Further experiments to search for structural transitions are in progress. It is considered, however, that accompanying changes in diffraction patterns can be very small, as in the case of $Cd_2Re_2O_7$, where the space group changes from cubic $Fd$-$3m$ to tetragonal $I$-$4m2$ at 200 K with negligible tetragonal distortion.[36]

## 5. Conclusions

In summary, we have performed resistivity measurements on a high-quality single crystal of $KOs_2O_6$ and powder X-ray diffraction experiments under HP up to 5 GPa. A first-order phase transition from a cubic phase with high $T_c$s to a lower-symmetry phase with low $T_c$s is observed at $P_s$ = 3.6 GPa. A large reduction in $T_c$ from 6.5 to 3.3 K at $P_s$ suggests that the increase in $T_c$ by the rattling has been ab-rogated above $P_s$. Two anomalies in resistivity are observed below $P_s$, which suggests that there are more structural transitions accompanying minor changes in rattling vibration.

**Acknowledgments**

The HP diffraction experiments were carried out under the approval of JASRI (Proposal Nos. 2007A2089 and 2007B1183). This work was partly supported by Grant-in-Aids for Scientific Research B (22340092), Scientific Research C (20540311), Scientific Research on Priority Areas (18027017 and 19052003), and Scientific Research on Innovative Areas (20102005) provided by MEXT, Japan.

1) S. Yonezawa, Y. Muraoka, Y. Matsushita, and Z. Hiroi: J. Phys.: Condens. Matter **16** (2004) L9.
2) M. Brühwiler, S. M. Kazakov, J. Karpinski, and B. Batlogg: Phys. Rev. B **73** (2006) 094518.
3) Z. Hiroi, S. Yonezawa, Y. Nagao, and J. Yamaura: Phys. Rev. B **76** (2007) 014523.
4) Y. Nagao, J. Yamaura, H. Ogusu, Y. Okamoto, and Z. Hiroi: J. Phys. Soc. Jpn. **78** (2009) 064702.
5) W. A. Harrison, *Solid State Theory* (McGraw-Hill Book Company, New York, 1970) 389.
6) A. D. Caplin and L. K. Nicholson: J. Phys. F **8** (1978) 51.
7) J. Yamaura, S. Yonezawa, Y. Muraoka, and Z. Hiroi: J. Solid State Chem. **179** (2006) 336.
8) J. Kunes, T. Jeong, and W. E. Pickett: Phys. Rev. B **70** (2004) 174510.
9) R. Galati, C. Simon, P. F. Henry, and M. T. Weller: Phys. Rev. B **77** (2008) 104523.
10) Z. Hiroi, S. Yonezawa, T. Muramatsu, J. Yamaura, and Y. Muraoka: J. Phys. Soc. Jpn. **74** (2005) 1255.
11) K. Sasai, K. Hirota, Y. Nagao, S. Yonezawa, and Z. Hiroi: J. Phys. Soc. Jpn. **76** (2007) 104603.
12) H. Mutka, M. M. Koza, M. R. Johnson, Z. Hiroi, J. Yamaura, and Y. Nagao: Phys. Rev. B **78** (2008) 104307.
13) T. Hasegawa, Y. Takasu, N. Ogita, M. Udagawa, J. Yamaura, Y. Nagao, and Z. Hiroi: Phys. Rev. B **77** (2008) 064303.
14) J. Schoenes, A.-M. Racu, K. Doll, Z. Bukowski, and J. Karpinski: Phys. Rev. B **77** (2008) 134515.
15) T. Dahm and K. Ueda: Phys. Rev. Lett. **99** (2007) 187003.
16) M. Yoshida, K. Arai, R. Kaido, M. Takigawa, S. Yonezawa, Y. Muraoka, and Z. Hiroi: Phys. Rev. Lett. **98** (2007) 197002.
17) Z. Hiroi, J. Yamaura, S. Yonezawa, and H. Harima: Physica C **460-462** (2007) 20.
18) T. Muramatsu, N. Takeshita, C. Terakura, H. Takagi, Y. Tokura, S. Yonezawa, Y. Muraoka, and Z. Hiroi: Phys. Rev. Lett. **95** (2005) 167004.
19) K. Miyoshi, Y. Takaichi, Y. Takamatsu, M. Miura, and J. Takeuchi: J. Phys. Soc. Jpn. **77** (2008) 043704.
20) R. Saniz and A. J. Freeman: Phys. Rev. B **72** (2005) 024522.
21) T. Isono, D. Iguchi, Y. Machida, K. Izawa, B. Salce, J. Flouquet, H. Ogusu, J. Yamaura, and Z. Hiroi: Physica C, in press.
22) S. Katrych, Q. F. Gu, Z. Bukowski, N. D. Zhigadlo, G. Krauss, and J. Karpinski: J. Solid State Chem. **182** (2009) 428.
23) N. Mori, C. Murayama, H. Takahashi, H. Kaneko, K. Kawabata, Y. Iye, S. Uchida, H. Takagi, Y. Tokura, Y. Kubo, H. Sasakura, and K. Yamaya: Physica C **185-189** (1991) 40.
24) K. Murata, K. Yokogawa, H. Yoshino, S. Klotz, P. Munsch, A. Irizawa, M. Nishiyama, K. Iizuka, T. Nanba, T. Okada, Y. Shiraga, and S. Aoyama: Rev. Sci. Instrum. **79** (2008) 085101.
25) Z. Hiroi, T. Yamauchi, T. Yamada, M. Hanawa, Y. Ohishi, O. Shimomura, M. Abliz, M. Hedo, and Y. Uwatoko: J. Phys. Soc. Jpn. **71** (2002) 1553.
26) R. Saniz, J. E. Medvedeva, L. H. Ye, T. Shishidou, and A. J. Freeman: Phys. Rev. B **70** (2004) 100505(R).
27) H. Harima: J. Phys. Chem. Solids **63** (2002) 1035.
28) D. Errandonea, Y. Meng, M. Somayazulu, and D. Häusermann: Physica B **355** (2005) 116.
29) M. A. Subramanian, G. Aravamudan, and G. V. S. Rao: Prog. Solid State Chem. **15** (1983) 55.
30) K. Hattori and H. Tsunetsugu: Phys. Rev. B **81** (2010) 134503.
31) Z. Hiroi, S. Yonezawa, and J. Yamaura: J. Phys.: Condens. Matter **19** (2007) 145283.
32) J. Yamaura, M. Takigawa, O. Yamamuro, and Z. Hiroi: J. Phys. Soc. Jpn. **79** (2010) 043601.
33) K. Sasai, M. Kofu, R. M. Ibberson, K. Hirota, J. Yamaura, Z. Hiroi, and O. Yamamuro: J. Phys.: Condens. Matter **22** (2010) 015403.
34) K. Hattori and H. Tsunetsugu: J. Phys. Soc. Jpn. **78** (2009) 013603.
35) K. Umeo, H. Kubo, J. Yamaura, Z. Hiroi, and T. Takabatake: J. Phys. Soc. Jpn. **78** (2009) 123602.
36) J. Yamaura and Z. Hiroi: J. Phys. Soc. Jpn. **71** (2002) 2598.